\documentclass{pasj00}

\SetRunningHead{H.\ Shibai et al.}{Absorption Measurement of Hydrogen Molecules}

\title{Absorption Measurement of Hydrogen Molecules in the Early Universe}

\author{Hiroshi \textsc{Shibai}, Tsutomu T. \textsc{Takeuchi}, 
  T. N. \textsc{Rengarajan}\thanks{Invitation Fellow of the Japan Society for
    the Promotion of Science.}}
\affil{Graduate School of Science, Nagoya University,
  Chikusa-ku, Nagoya 464--8602}
\email{shibai@phys.nagoya-u.ac.jp}
\and
\author{Hiroyuki \textsc{Hirashita}\thanks{Research Fellow of the Japan Society for
    the Promotion of Science.}}
\affil{Department of Astronomy, Kyoto University, Sakyo-ku, Kyoto 606--8502}

\Received{$\langle$reception date$\rangle$}
\Accepted{$\langle$acception date$\rangle$}
\Published{$\langle$publication date$\rangle$}

\begin{document}
\maketitle

\begin{abstract}
We investigate the observability of hydrogen molecules in absorption.
The absorption efficiency of the hydrogen molecules become comparable with 
or larger than that of the dust grains in the metal-poor condition  
expected in the early Universe.
If we can use bright infrared continuum sources behind the molecular gas 
clouds, the absorption measurement of the hydrogen molecules will be an 
important technique to explore the primordial gas clouds that are contracting 
into first-generation objects. 
\end{abstract}

\section{Introduction}

Molecular hydrogen is the predominant constituent of the dense gas in the 
Universe. 
One of the most important questions in current astrophysics is how the first 
generation objects formed from the primordial gas that was not polluted by 
heavy elements (e.g. Omukai \& Nishi 1998). 
Theoretical models should be confirmed by observations to establish a more 
realistic scenario of the first generation objects in the Universe. 


\begin{figure*}[bt]
\begin{center}
\FigureFile(170mm,127mm){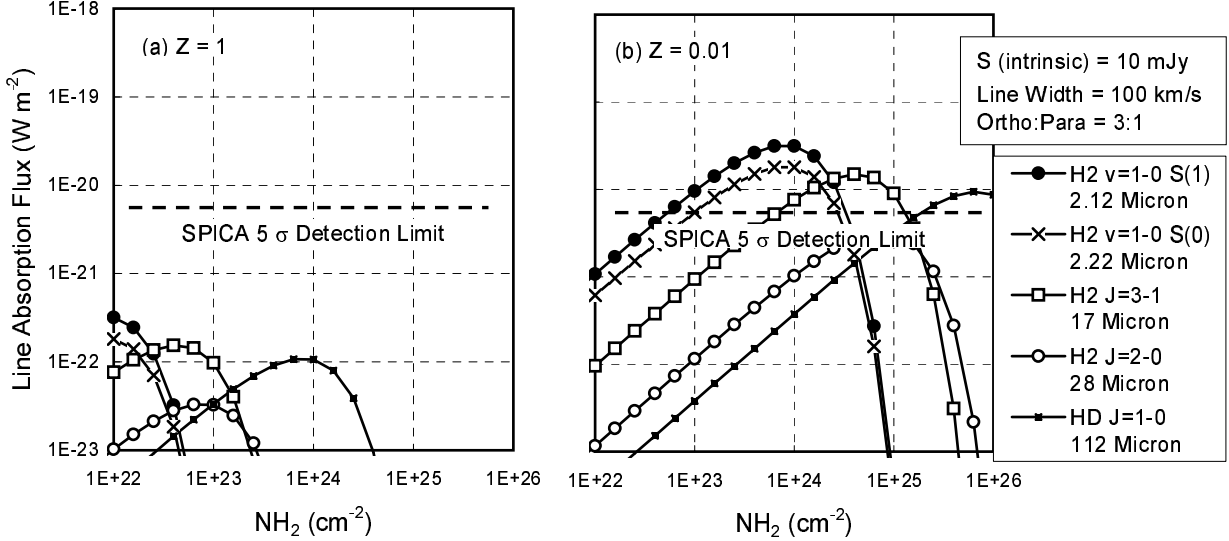}
\end{center}
\vspace*{-45mm}
\caption{$(a)$ Absorption line flux expected for a cloud with the Milky Way heavy 
  element abundance ($Z = 1$) in front of a 10~mJy source.
  $(b)$ Same as $(a)$ but the heavy element abundance is $Z = 0.01$.}
\end{figure*}


\begin{table*}
\begin{center}
\caption{Transitions of molecular hydrogen.}
\label{tab1}
\begin{tabular}{lll} \hline \hline
Transition & Wavelength  & $A$ coefficients \\
\hline
ionization and dissociation & UV & large \\
vib-rotational transitions  & NIR (2 $\mu$m) & small \\
rotational transitions  & MIR (28, 17, $\cdots \;\mu$m) & small\\
rotational transitions of HD & FIR (112, 56, $\cdots \;\mu$m) & intermediate \\
\hline
\end{tabular}
\end{center}
\end{table*}

While in the local Universe, molecules containing heavy elements (e.g.\
CO, ${\rm H_2O}$) are good tracers of the amount of molecular hydrogen, 
they are expected to be significantly depleted in the early Universe,
though some CO emission lines have been detected for high-$z$ objects (e.g.\ 
Ohta et al.\ 1996).
We need a technique for measuring molecular hydrogen directly. 
Petitjean et al.\ (2000) have reported a direct measurement of ${\rm H}_2$
molecules in the ultraviolet.
Table~\ref{tab1} describes some important transitions of hydrogen molecules. 
The transition probabilities of ionization and dissociation lines are 
so large that they are useful for detecting thin layers and 
small amounts of the molecular gas, but useless for detecting dense gas clouds.
On the other hand, molecular hydrogen has well-known vibrational and 
rotational transitions in the infrared (IR) region.
Their transition probabilities are very small because hydrogen molecule, 
a diatomic molecule of two identical nuclei, has no allowed dipole 
transitions but has allowed quadrupole transitions. 

These transitions of the hydrogen molecule can have both emission and 
absorption. 
Vib-rotational and rotational line emission from hydrogen molecules are 
useful tools to analyze dense 
($> 10\; {\rm cm}^{-3}$) and hot ($> 300$~K) gas. 
Expected emission line intensities of molecular hydrogen were calculated by 
Ciardi \& Ferrara (2001). 
According to their result, direct measurement of the emission lines of 
molecular hydrogen is very difficult due to their weakness. 
On the other hand, if there is a strong IR continuum source behind or 
in the molecular gas cloud, absorption measurements of these transition 
lines may be possible.
The absorption measurement has the great advantage that the column density 
of the absorber can be derived with fewer assumptions. 

However, we should note that the absorption by dust in the cloud is 
usually larger than the absorption by ${\rm H}_2$ molecules. 
If the absorber is metal rich as in the ISM of the Galaxy, the extinction
by dust causes the emergent continuum flux to decrease fast, thus making it
difficult to observe small absorption signals (Lacy et al.\ 1994).
Fortunately, metal abundance is expected to be significantly 
lower in earlier epochs of the Universe; hence the dust absorption is 
considerably less than that in the Galaxy. 
Therefore, hydrogen molecular lines can be detected in absorption against 
bright IR sources more easily in the early Universe than in the Galaxy. 
Such observation will be feasible with the advent of proposed space missions 
for large IR telescope facilities, e.g.\ {\sl SPICA} ({\sl HII/L2}$\,$; 
Nakagawa et al.\ 2000), {\sl SPIRIT} (Mather 2000).

In this paper we investigate this possibility and consider what 
we can learn from such observations.
The rest of the manuscript is as follows:
In Section~2 we present the formulation of the intensity of absorption by 
hydrogen molecules.
Our results are shown in Section~3.
We discuss some relevant issues in Section~4.

\section{Calculation}

Assuming a uniform, cool gas cloud ($kT_{\rm ex} \ll h\nu$), the optical 
thickness of the line absorption, $\tau_{\rm line}$ is 
\begin{equation}
  \tau_{\rm line} \simeq \frac{\lambda^3}{8\pi} \left(
    \frac{g_{\rm u}}{g_{\rm l}} \right) A_{\rm ul} N \frac{1}{\Delta v} \;,
\end{equation}
where the subscripts u and l indicate the upper and lower levels of a 
transition, $g_{\rm u}$ and $g_{\rm l}$ are the degeneracy of each 
state respectively, $A_{\rm ul}$ is the Einstein's $A$~coefficient, 
$N$ is the column density of the molecules in the lower state, 
and $\Delta v$ is the line width in units of velocity. 
Here we assume that almost all the molecules occupy the lowest energy state.
As is well known, in the optically thin case, the column density can 
directly be derived from the equivalent width, which is nearly equal to the 
product of $\tau_{\rm line}$ and $\Delta v$.   

The absorption line flux in the extinction free case, $I^{\rm abs}_{\rm 
line,0}$, is obtained by
\begin{equation}
  I^{\rm abs}_{\rm line,0} = S \Delta \nu 
  \left( 1 - \exp (-\tau_{\rm line}) \right) \;,
\end{equation}
where $\Delta \nu$ is the line width in units of frequency and $S$ is 
the continuum flux of the IR source behind the cloud.
The line width and the source flux are assumed to be $100\; {\rm km\,s^{-1}}$ 
and 10~mJy, respectively. 
We can scale the results of this calculation for other source fluxes, for 
example, based on the capability of the experimental set up.
Assumed parameters are listed in Table~\ref{tab2}.


\begin{table*}
\begin{center}
\caption{Parameters assumed for the present calculation.}
\label{tab2}
\begin{tabular}{cc} \hline \hline
Ortho : Para & 3 : 1 \\
HD/${\rm H}_2$ & $10^{-5}$ \\
Intrinsic Flux of Source & 10~mJy \\
Line width & 100 ${\rm km\,s}^{-1}$ \\
Detection Limit (5$\sigma$) $^{a}$ & $5\times 10^{-21}\; {\rm W\,m}^{-2}$\\
Detectable Optical Thickness & $> 0.01$ \\
Dust Extinction Model & Mathis (1990) \\
\hline
$a$:  Expected detection limit of the {\sl SPICA} mission 
(Ueno et al.\ 2000).\\
\end{tabular}
\end{center}
\end{table*} 

Next, we consider about the dust extinction which is denoted by
\begin{equation}\label{eq:tau_dust}
  \tau_{\rm dust} = 1.087 \left( \frac{A_\lambda}{A_V} \right) 
  \left( \frac{A_V}{N_{\rm H}} \right)_{\odot} Z \, N_{\rm H} \; .
\end{equation}
The extinction spectrum of Mathis (1990) is adopted for 
($A_{\lambda}$/$A_V$), and the extinction efficiency is assumed to be 
proportional to the relative heavy element abundance, 
and $(A_V/N_{\rm H})_{\odot}$ is the conversion factor from $A_V$ to 
$N_{\rm H}$ for the local abundance listed in Table~\ref{tab2}.

Finally, we obtain the absorption line flux with extinction, 
$I^{\rm abs}_{\rm line}$, as
\begin{eqnarray}
  I^{\rm abs}_{\rm line} &=& I^{\rm abs}_{\rm line,0}
  \exp(-\tau_{\rm dust}) \nonumber \\
  &=& S  \Delta \nu  (1- \exp(-\tau_{\rm line})) \exp(-\tau_{\rm dust}) \;.
\end{eqnarray}

\section{Results}


\begin{table*}
\begin{center}
\caption{Parameters of the Line Transitions.}
\label{tab3}
\begin{tabular}{ccllc} \hline \hline
Transition & Wavelength [$\mu$m]  & $A$ Coefficient [${\rm s}^{-1}$] & 
${\cal N}_{\rm H_2}\,[\mbox{cm}^{-2}]$ for $\tau_{\rm line} = 0.01$ & 
$A_{\lambda}/A_V$\\
\hline
${\rm H_2}\; v = 1-0$ $S(1)$ & 2.12 & $3.47 \times 10^{-7}$ & 
$4 \times 10^{23}$ & 0.11\\
${\rm H_2}\; v = 1-0$ $S(0)$ & 2.22 & $2.53 \times 10^{-7}$ & 
$8 \times 10^{23}$ & 0.11\\
${\rm H_2}\; v = 0-0$ $S(1)$ & 17 & $4.77 \times 10^{-10}$ & 
$7 \times 10^{23}$ & 0.020\\
${\rm H_2}\; v = 0-0$ $S(0)$ & 28 & $2.95 \times 10^{-11}$ & 
$3 \times 10^{24}$ & 0.011\\
HD $v = 0-0$ $R(0)$ & 112 & $2.54 \times 10^{-8}$ & 
$2.5 \times 10^{24}$ & 0.0011\\
\hline
\end{tabular}
\end{center}
\end{table*} 

The calculation was made for the five lines listed in Table~3. 
Parameters used here are from Turner et al.\ (1977). 
First, it is interesting to compare the optical thickness of the line 
absorption and the dust extinction. 
In case of $\Delta v = 100 \;{\rm km\,s^{-1}}$ and the local heavy element 
abundance, the ratio of the line optical thickness to the dust optical 
thickness is $10^{-4}$ for vib-rotational lines in NIR, $10^{-3}$ for pure 
rotational lines in MIR, and $10^{-2}$ for pure rotational lines of HD in 
the FIR. 
It means that the absorption measurement is certainly difficult against the 
large extinction in the absorbing cloud. 
However, this ratio increases with the inverse of the heavy element abundance. 
Therefore, in the lower heavy element abundance case, we can expect 
reasonably higher ratios.

It is quite difficult to detect absorption lines whose optical thickness 
is less than 1~\%. 
Therefore, as seen from Table~2, only those clouds whose column density 
is larger than $10^{24}\; {\rm cm}^{-2}$ can be detected in absorption.

Figure~1a shows the absorption line flux expected for a cloud with 
the local heavy element abundance in front of a 10~mJy source. 
The absorption fluxes are far smaller than the $5 \sigma $ expected 
sensitivity of the {\sl SPICA} mission (Ueno et al.\ 2000). 
On the other hand, Figure~1b shows the result for the case in which 
the heavy element abundance is 1~\% of that of the local one. 
All five lines populate parameter space above or near the limit.

Figure~2 shows the result of the same calculation, but for the 
${\rm H}_2$ $v = 0-1$ $S(1)$ line for various values of the heavy element 
abundance. 
In case of $N_{\rm H_2} > 10^{24}\; {\rm cm}^{-2}$ and $Z < 0.01$, 
the absorption line can be detected.  

\section{Discussion}

\subsection{Possible Energy Source}

In this subsection we discuss possible energy sources at high redshifts.
In our calculations, we have assumed an intrinsic continuum flux density 
of 10~mJy at the line wavelength.
What sort of flux density can we expect from high-$z$ objects? To
investigate this, we have plotted in Figure~3 the SED in the rest system 
of several high redshift objects for which observations are available.
The sources and references for them are as follows: IRAS~F$10214+4724$ 
($z = 2.286$; Rowan-Robinson et al.\ 1993; Barvainis et al.\ 1995), 
H$1413+117$ ($z = 2.558$; Barvainis et al.\ 1995), 
SMM~$02399-0136$ ($z= 2.8$; Ivison et al.~1998), 
SMM~J$14011+0252$ ($z= 2.55$; Ivison et al.~2000), 
APM~$08279+5255$ ($z=3.87$; Lewis et al.~1988), 
and BR~$1202-0725$ ($z = 4.69$; Isaak et al.~1994).
In this plot we have not corrected for the possible lens amplification for 
the first three objects and all observations have been transformed to a 
common redshift of 5 ($q_0 = 0.1$).
The rest system SEDs and flux densities of F$10214+4724$ and H$1413+117$
are almost identical and encompass three of the lines -- 17, 28, and 
112~$\mu$m.
The interpolated flux densities at the three frequencies are 8, 16, and 
9~mJy respectively.
For the rest of the sources, we extrapolate from the observed 450~$\mu$m
flux density assuming the rest system SED to be the same as that of 
F$10214+4724$. From
this exercise we find the range of flux densities to be $1.6\mbox{--}19$~mJy
at 17~$\mu$m, $3\mbox{--}37$~mJy at 28~$\mu$m, and $2\mbox{--}21$~mJy at 
112~$\mu$m.
On the other hand, if we use the SED of Arp~220, the expected flux densities 
are about 38 times larger at 17~$\mu$m, 11 times at 28~$\mu$m, and about
the same at 112~$\mu$m.
For distant highly luminous sources, Arp~220 SED may not be appropriate since
Solomon et al.\ (1997) find that for a sample of ultraluminous IR 
galaxies out to a redshift of 0.3, the emission at 100~$\mu$m may be optically thick.
Following their finding, if we use a 60~K blackbody spectrum as a template,
the expected flux densities at 17 and 28~$\mu$m are a factor of 5 and 3 lower 
as compared with using Arp~220 SED.

Thus, if we have objects at $z = 5$ at least as luminous as our template 
sources, it is possible to observe absorption in ${\rm H_2}$ lines of 17, 28,
and 112~$\mu$m respectively.
The redshifted 112~$\mu$m line will be in the submm range and may not be 
observable by space missions of near future.
However, the proposed large area and high sensitivity ground-based 
mm--submm interferometric arrays like ALMA should be able to observe it.
Unfortunately, it is very difficult to observe the absorption of 2.2~$\mu$m
lines since the expected flux density based on dust emission spectrum is a 
factor of 10 or more lower as compared with that at 17~$\mu$m.
However, if a bright non-thermal flat spectrum source is present, absorption 
measurement at 2.2~$\mu$m may be possible.

\subsection{What can we learn from the absorption lines of primordial clouds?}


\begin{figure}[t]
\begin{center}
\FigureFile(80mm,80mm){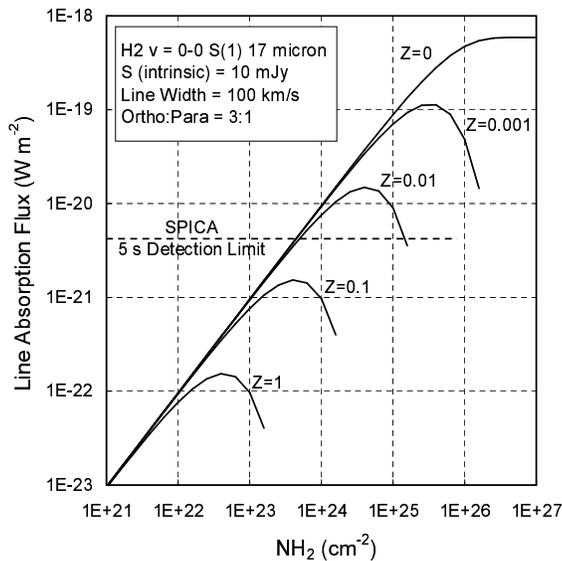}
\end{center}
\caption{Result of the same calculation as in Fig.~1, but for the $\mbox{H}_2$ 
  $v = 0-0$ $S(1)$ line in various values of the heavy element abundance.}
\end{figure}

First we simply estimate the column density of a protogalactic 
hydrogen cloud.
Here we assume that the size of the cloud $R$ is a few kpc 
and that the gas mass $M$ is $\sim 10^{11}\;M_\odot$.
Such a large reservoir of molecular gas has been discovered at high redshift 
(Papadopoulos et al.\ 2001). 
Such clouds have column density of
\begin{eqnarray}
  N_{\rm H_2} 
  \simeq 4 \times 10^{23} \; [{\rm cm}^{-2}] \, f 
  \left(\frac{R}{3\;[{\rm kpc}]}\right)^{-2}
  \left( \frac{M}{10^{11}M_\odot}\right) \; .
\end{eqnarray}
where $f$ is the mass fraction of the molecular clouds to the total gas mass,
and $M$ is the total gas mass of the protogalaxy.
The final fate of such a large reservoir of gas may be the formation
of a giant galaxy, whose typical core radius is a few kpc. 
If such a collapse really occurs in the formation epoch of galaxies,
resulting column density is as large as the minimum column density 
($7 \times 10^{23}\;{\rm cm}^{-2}$) needed for detecting
${\rm H_2}$ line absorption.


\begin{figure}[t]
\begin{center}
\FigureFile(80mm,80mm){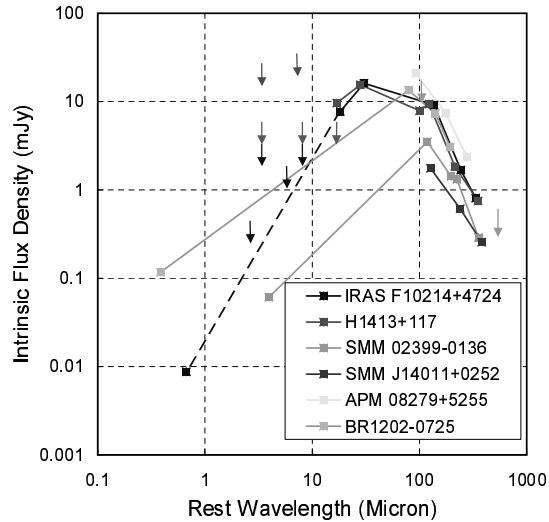}
\end{center}
\caption{The spectral energy distributions of six representative 
  high-$z$ objects.
  Downward arrows represent the upper limits.}
\end{figure}

In a realistic situation, a primordial cloud may evolve dynamically on
a free-fall timescale.
The free-fall timescale is much shorter than the timescale of cosmological 
structure evolution, e.g.\ the Hubble timescale.
Therefore, observed properties are specific to the redshift at which the
cloud absorption is measured.
We will obtain redshift $z$, and velocity dispersion $\Delta v$.
The present structure formation theory provides us the relation between 
the velocity dispersion and the corresponding dynamical mass of the 
cloud as a function of redshift, $M_{\rm vir}(z)$ (cf. Padmanabhan 1993).
Thus, we can constrain the structure formation theory.
The SPICA mission will provide us with the information on the massive
objects ($\gtsim 10^{11}~\MO$) at $z\ltsim 5$. 
We still have to wait for more sensitive facilities for the observational
approach to the ``Population III'' objects, whose physical properties 
are theoretically predicted by recent extensive investigations, 
because they are located at higher redshift and their typical mass is 
small (e.g., Nishi \& Susa 1999).
When we estimate the detectability of such objects by other facilities, 
the algorithm constructed in this paper can be applied straightforwardly.

The above estimate (equation 5) is based on the mean column
density of a galaxy. If we have a chance to detect the final
phase of a collpsing gas in the line of sight of a quasar,
we can constrain the physical state of such a collpase. In other
words, we
will be able to measure the column density $N_{\rm H_2}$, 
and excitation temperature $T_{\rm ex}$, which are interesting quantities
in the context of the evolution of an individual cloud.
Recent theoretical work (e.g., Omukai \& Nishi 1998; Omukai 2000) proposes 
the following scenario. 
The collapsing primordial gas must radiate its gravitational energy before 
the first generation objects are born from it. 
However, the gas consisting of only hydrogen and helium atoms cannot cool 
efficiently below $\mbox{a few} \times 10^3$ K because their constituents 
do not have radiative transitions corresponding $\mbox{a few} \times 10$ K
to $\mbox{a few} \times 10^3$ K. 
Therefore, a much more efficient cooling process must be working at this 
stage of the Universe. 
The most plausible cooling process is vib-rotational and pure rotational 
lines of molecular hydrogen. 
If the molecular hydrogen is effectively produced in the collapsing 
primordial gas at $\mbox{a few} \times 10^3$ K, the expected time scale of 
the gas contraction is considerably reduced. 
Galli \& Palla (1998) indicated that so called relic electrons would work 
there as effective catalyst to produce primordial molecular clouds.
If we have a high-precision measurements and follow-up observations of the
absorption cloud, we might be able to pursue the dynamical evolution of the
cloud and to compare the theoretical predictions.

\vspace{1pc}\par
The authors thank H. Matsuhara, T. Nakagawa, and M. Harwit for their useful comments.
TTT and HH are grateful to B. Ciardi and A. Ferrara for stimulating 
discussions with them.
TNR acknowledges the Invitation Fellowships of the Japan Society for 
the Promotion of Science.
HH acknowledges the Research Fellowships of the Japan Society for 
the Promotion of Science for Young Scientists.

\section*{References} \vspace{1mm}

\re
Barvainis, R., Antonucci, R., Hurt, T., Coleman, P., \& Reuter, H.-P.\ 1995, 
\apj , 451, L9

\re 
Ciardi, B., \& Ferrara, A. 2001, \mnras , in press, astro-ph/0005461

\re 
Galli, D., \& Palla, F. 1998, \aap , 335, 403

\re
Isaak, K.\ G., McMahon, R.\ G., Hills, R.\ E., \& Withington, S.\ 
1994, \mnras , 269, L28

\re
Ivison, R. J., Smail, I., Le Borgne, J.-F., Blain, A. W., Kneib, J.-P., 
Bezecourt, J., Kerr, T. H., \& Davies, J. K. 1998, \mnras , 298, 583

\re
Ivison, R. J., Smail, I., Barger, A. J., Kneib, J.-P., Blain, A. W., 
Owen, F. N., Kerr, T. H., \& Cowie, L. L.\ 2000, \mnras , 315, 209  

\re
Klaas, U., Haas, M., Heinrichsen, I., \& Schulz, B. 1997, \aap , 325, L21

\re 
Lacy, J. H., Knacke, R., Geballe, T. R., Tokunaga, A. T. 1994, \apjl , 428, L69

\re
Lewis, G.\ F., Chapman, S.\ C., Ibata, R.\ A., Irwin, M.\ J., \& Totten, E.\ J.
1998, \apjl , 505, L1

\re
Mather, J.\ 2000, in `Mid- and Far-Infrared Astronomy and Future Space 
Missions', ed T.\ Matsumoto \& H.\ Shibai, ISAS Report, SP-14, p219

\re 
Mathis, J.\ S.\ 1990, \araa , 28, 37

\re 
Nakagawa, T., and the HII/L2 Mission Working Group 2000, 
in `Mid- and Far-Infrared Astronomy and Future Space 
Missions', ed T.\ Matsumoto \& H.\ Shibai, ISAS Report, SP-14, p189

\re
Nishi, R., \& Susa, H. 1999, \apjl , 523, L103

\re
Ohta, K., Yamada, T., Nakanishi, K., Kohno, K., Akiyama, M., \& Kawabe, R.
1996, Nature, 382, 426

\re 
Omukai, K. \& Nishi, R. 1998, \apj , 508, 141

\re 
Omukai, K.\ 2000, \apj , 534, 809

\re
Papadopoulos, P., Ivison, R., Carilli, C., \& Lewis, G.\ 2001, Nature, 
409, 58

\re
Petitjean, P., Srianand, R., \& Ledoux, C. 2000, \aap , 364, L26

\re
Rowan-Robinson, M., Eftasthiou, A., Lawrence, A., Oliver, S., Taylor, A., 
Broadhurst, T. J., McMahon, R. G., Benn C. R., et al.\ 1993, \mnras , 231, 513  

\re
Solomon, P.\ M., Downes, D., Radford, S.\ J.\ E., \& Barrett, J.\ W.
1997, \apj , 478, 144

\re 
Turner, J., Kirby-Docken, K., \& Dalgarno, A. 1977, \apjs , 35, 281 

\re 
Ueno, M., Matsuhara, H., Murakami, H., \& Nakagawa, T.\ 2000, 
in `Mid- and Far-Infrared Astronomy and Future Space 
Missions', ed T.\ Matsumoto \& H.\ Shibai, ISAS Report, SP-14, p197

\label{last}

\end{document}